\documentclass[a4paper]{article}
\usepackage{amsmath,graphicx,amsfonts,breqn,color,
mathtools,amssymb,subcaption,booktabs,bm,easybmat}
\usepackage{INTERSPEECH2019}

\def\vec#1{\ensuremath{\bm{{#1}}}}
\def\mat#1{\mathbf#1}
\newcommand{\tr}{^{\mathsf T}}

\newcommand{\mbf}{\mathbf}
\newcommand{\bT}{\mat{T}}
\newcommand{\bS}{\mat{\Sigma}}

\title{Towards Adapting NMF Dictionaries Using Total Variability Modeling for Noise-Robust Acoustic Features}
\name{Kunal Dhawan$^{1}$\sthanks{The first author performed the work while at University of Southern California.}, Colin Vaz$^{2}$, Ruchir Travadi$^{2}$, and Shrikanth Narayanan$^{2}$}

\address{
  $^{1}$Indian Institute of Technology Guwahati, Guwahati, India 781039 \\
  $^{2}$Signal Analysis and Interpretation Lab, University of Southern California, Los Angeles, CA 90089
}
\email{k.dhawan@iitg.ac.in, <cvaz,travadi>@usc.edu shri@sipi.usc.edu}

\begin{document}
%
\maketitle
\begin{abstract}
We propose an algorithm to extract noise-robust acoustic features from noisy speech. We use Total Variability Modeling in combination with Non-negative Matrix Factorization (NMF) to learn a total variability subspace and adapt NMF dictionaries for each utterance. Unlike several other approaches for extracting noise-robust features, our algorithm does not require a training corpus of parallel clean and noisy speech. Furthermore, the proposed features are produced by an utterance-specific transform, allowing the features to be robust to the noise occurring in each utterance. Preliminary results on the Aurora 4 + DEMAND noise corpus show that our proposed features perform comparably to baseline acoustic features, including features calculated from a convolutive NMF (CNMF) model. Moreover, on unseen noises, our proposed features gives the most similar word error rate to clean speech compared to the baseline features.
\let\thefootnote\relax\footnotetext{The authors would like to acknowledge the support of NIH grant R01DC007124, NSF grant 1514544, and the IUSSTF-Viterbi program.}
\end{abstract}
\section{Introduction}
\label{sec:intro}

Automatic speech recognition (ASR) systems are being increasingly deployed on a wide range of devices for a wide range of applications. Speech offers a natural and efficient way to interact with these devices. Furthermore, speech contains paralinguistic content that devices can use to modify their outputs or behavior. For example, ASR systems in call centers can use the emotion of customers to better serve them or mitigate conflicts \cite{pappas}. Given the wide usage scenarios, ASR systems need to perform robustly in different acoustic environments, with various background noises and channel conditions, and reliably recognize speech with different dialects and accents. Thus, there has been increasing research in making ASR systems more robust to various real-world conditions. Some techniques researchers have developed include speech denoising \cite{macho}, feature enhancement \cite{yoshioka}, feature transformation \cite{droppo}, and acoustic model adaptation \cite{kalinli,wang}.

Speech denoising is one straightforward way to make ASRs robust to background noise. Pre-processing the speech with a noise-removal algorithm reduces the mismatch between features extracted at test time compared to the features used to train the ASR. Common speech denoising algorithms include Weiner filtering and spectral subtraction \cite{boll}. Non-negative matrix factorization (NMF) \cite{paatero,leeseung} is also widely used for denoising. The drawback with speech denoising is that it usually introduces distortion and artifacts, such as musical noise, and has been shown to degrade ASR performance \cite{seltzer,narayanan}. Moreover, the artifacts are usually amplified when the background noise is highly non-stationary or energetic.

To overcome the drawbacks of speech denoising, researchers have investigated extracting acoustic features directly from noisy speech that are robust to noise.
Moreno et al. introduced Vector Taylor Series (VTS) features \cite{vts}, which uses the Taylor series expansion of the noisy signal to model the effect of noise and channel characteristics on the speech statistics.
Deng et al. proposed the Stereo-based Piecewise Linear Compensation for Environments (SPLICE) algorithm \cite{deng} for generating noise-robust features by assuming that clean speech cepstral vectors have a piece-wise linear relationship to noisy speech cepstral vectors. Power-Normalized Cepstral Coefficients (PNCC) \cite{pncc} draw inspiration from human auditory processing for generating noise-robust features, and were shown to reduce word error rates on noisy speech compared to Mel-Frequency Cepstral Coefficients (MFCC) and Relative Spectral Perceptual Linear Prediction (RASTA-PLP) coefficients. Recently, an NMF-based approach was proposed \cite{vaz}, where speech and noise dictionaries are trained on clean and noisy speech, and the coefficients in terms of these dictionaries are used as acoustic features. They showed that the NMF-based approach gives better ASR performance than log-mel features or denoising the speech.

In this work, we expand upon the NMF-based approach. On a training set, we learn a universal background model (UBM) dictionary, and then use total variability modeling \cite{dehak} to learn a subspace for adapting the UBM dictionary to different noise and channel conditions. The advantages of this proposed method over the one in \cite{vaz} are two-fold:
\begin{enumerate}
    \item The training set does not require parallel clean and noisy utterances, and
    \item The dictionary can be adapted for each utterance at test time, allowing for better modeling of the acoustic conditions in each utterance.
\end{enumerate}
In the following sections, we provide a brief overview of NMF and total variability modeling, followed by our proposed noise-robust acoustic feature algorithm. Section~\ref{sec:experiments} describes our experiments and offers insights into the results, and Section~\ref{sec:conclusion} offers our concluding remarks and future directions.

\section{Background}
\label{sec:background}

\subsection{Non-negative Matrix Factorization}
\label{sec:nmf_background}

NMF decomposes a non-negative matrix $\mat{V} \in \mathbb{R}_+^{d \times t}$ into the product of a non-negative dictionary $\mat{W} \in \mathbb{R}_+^{d \times k}$ and non-negative activation matrix $\mat{H} \in \mathbb{R}_+^{k \times t}$. Because of the non-negative constraint, the decomposition is purely additive, and one can think of the dictionary as containing $k$ components that are added together by the activation matrix to approximate the input matrix. In the case of speech, the input matrix is typically the magnitude spectrogram, and the dictionary contains spectral ``building blocks'' required to reconstruct the spectrogram.

\begin{figure}
\centering
\includegraphics[width=0.5\textwidth]{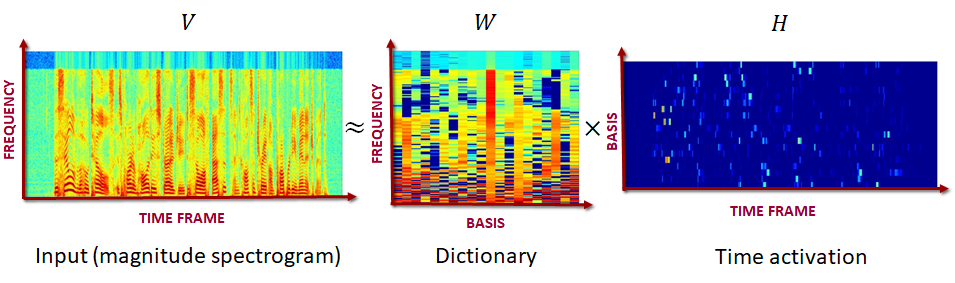}
\caption{Visualizing the dictionary $\mat{W}$ and activation matrix $\mat{H}$ after running NMF on a speech signal $\mat{V}$.}
\label{fig:visualization}
\end{figure}

For speech processing, it has been shown that the generalized KL divergence gives slightly better performance as the NMF cost function compared to the squared Euclidean distance \cite{fitzgerald}. Defining $\hat{\mat{V}} = \mat{W}\mat{H}$, the generalized KL divergence between $\mat{V}$ and $\hat{\mat{V}}$ is
\begin{equation}
\label{eq:kl_div}
    D_{GKL} \left( \mat{V} \| \hat{\mat{V}} \right) = \sum_{i=1}^{d} \sum_{j=1}^{t} \mat{V}_{ij} \ln \left( \frac{\mat{V}_{ij}}{\hat{\mat{V}}_{ij}} \right) - \mat{V}_{ij} + \hat{\mat{V}}_{ij},
\end{equation}
where $V_{ij}$ refers to the element in row $i$ and column $j$ of $\mat{V}$. Lee and Seung derived iterative multiplicative updates for $\mat{W}$ and $\mat{H}$ to minimize Equation~\ref{eq:kl_div} \cite{leeseung}. The advantages of using multiplicative updates over standard gradient descent updates are that no step size parameter is required and $\mat{W}$ and $\mat{H}$ are guaranteed to stay non-negative at each iteration if they are initialized with non-negative values. The NMF decomposition for the case of speech can be visualized in Figure~\ref{fig:visualization}.

\subsection{Total Variability Modeling}
\label{sec:tvm_background}

The Total Variability Model (TVM) \cite{dehak} is a tool which can be used to capture distributional differences between sequences of feature vectors within a fixed dimensional representation.
In particular, the assumption is that the feature vectors follow a distribution which has the form of a Gaussian Mixture Model (GMM) where the mean vectors corresponding to different Gaussian components vary across different utterances (in a constrained manner).

Let $\mbf{V} = \left\{ \mbf{V}_u\right\}_{u=1}^{U}$ be the collection of acoustic feature vectors in a dataset comprising $U$ utterances, where $\mbf{V}_u = \left\{ \vec{v}_{u \tau}\right\}_{\tau=1}^{t_u}$ denotes the feature vector sequence of length $t_u$ from a specific utterance $u$. Let $d$ be the dimensionality of each feature vector: $\vec{v}_{ut} \in \mathbb{R}^d$. 
  
It is assumed that with every utterance $u$, there is an associated vector $\vec{q}_u \in \mathbb{R}^s$, known as the \emph{ivector} for that utterance, such that the conditional distribution of $\vec{v}_{u \tau}$ given $\vec{q}_u$ is a GMM with $k$ components, and parameters $\left\{ p_c, \vec{\mu}_{uc} = \vec{\mu}_c + \mat{T}_c \vec{q}_u, \bS_c\right\}_{c=1}^{k}$
where $p_c\in\mathbb{R},\vec{\mu}_c\in\mathbb{R}^d, \mbf{T}_c \in \mathbb{R}^{d \times s} $ and $\mbf{\Sigma_c} \in \mathbb{R}^{d \times d}$. 
The prior distribution for $\vec{q}_u$ is assumed to be standard normal: 
\begin{equation} 
  \vec{q}_u \sim \mathcal{N}(\vec{0},\mbf{I})
\end{equation}
Let $\vec{m}_0, \vec{m}_u \in \mathbb{R}^{kd}$, known as supervectors, denote vectors consisting of stacked global and utterance-specific component means $\vec{\mu}_c$ and $\vec{\mu}_{uc}$ respectively. 
Then, TVM can be summarized as:
\begin{equation}
  \vec{M}_u = \vec{M}_0 + \mbf{T}\vec{q}_u
\end{equation}
where $\mbf{T} \in \mathbb{R}^{kd \times s}$ is given as:
$\bT = \left[ \begin{BMAT}(r){c.c.c}{c} \,\bT_1\tr\, & \,\dots\, & \,\bT_C\tr\, \end{BMAT} \right]\tr$

\section{Algorithm}
\label{sec:algorithm}

In this section, we describe an algorithm that uses TVM to adapt an NMF dictionary to the noise in an input spectrogram. The idea is for the dictionary to capture as much of the noise in the spectrogram as possible so that the activation matrix is not affected by noise. We will use the activation matrix as acoustic features for ASR on noisy speech. In our algorithm, the input features for the total variability model are magnitude spectrogram, the dictionary vectors play the role of GMM mean vectors, and the column-normalized activation matrix act as the posteriors of each GMM component. Table~\ref{tab:tvm_nmf_parallels} summarizes how we fit NMF in the TVM paradigm, and the following subsections describe the algorithm in detail. 


\begin{table}[h]
    \centering
    \caption{Connections between NMF matrices and TVM inputs.}
    \label{tab:tvm_nmf_parallels}
    \begin{tabular}{l|l}
        \hline
        \hline
         TVM Inputs & NMF Matrices \\
         \hline
         Features & Magnitude spectrogram \\
         GMM mean supervector & Vectorized NMF dictionary \\
         Cluster posteriors & Normalized activation matrix \\
         \hline
         \hline
    \end{tabular}
\end{table}


\subsection{Step 1: Learning UBM Dictionary}
\label{sec:nmf}
\vspace{-1mm}



We concatenate the utterances from the training set with various noisy conditions and compute the magnitude spectrogram $\mat{V}_{\text{all}}$. We use NMF to decompose $\mat{V}_{\text{all}} \in \mathbb{R}_+^{d \times t}$ into a $d \times k$ dictionary $\mat{W}_{\text{ubm}}$ and $k \times t$ activation matrix $\mat{H}_{\text{ubm}}$. We will refer to $\mat{W}_{\text{ubm}}$ as the UBM dictionary as it models the salient spectral components in the presence of all sources of variability. Thus, one can think of the $k$ vectors in the UBM dictionary as containing $k$ salient points in the feature space.


It has been reported in literature that incorporating a sparsity constraint on the activation matrix while applying NMF leads to a more expressive dictionary \cite{leroux,hoyer}. Thus, we add an $\ell_1$ penalty on the activation matrix to the generalized KL divergence cost function to encourage sparsity:
\begin{dmath}
\label{eq:costfunc}
C = D_{GKL} \left( \mat{V}_{\text{all}} \| \hat{\mat{V}}_{\text{all}} \right) + \lambda \sum_{i=1}^k \sum_{j=1}^t \mat{H}_{\text{ubm}(i,j)},
\end{dmath}
where $ \hat{\mat{V}}_{\text{all}} \vcentcolon =  \mat{W}_{\text{ubm}} {\mat{H}}_{\text{ubm}}$ and $\lambda$ controls the level of sparsity of $\mat{H}$. To minimize Equation~\ref{eq:costfunc}, we iteratively update $\mat{W}_{\text{ubm}}$ and $\mat{H}_{\text{ubm}}$ with the following multiplicative updates:
\begin{subequations}
\label{eq:speechupdates}
\begin{align}
\mat{W}_{\text{ubm}} & \leftarrow \mat{W}_{\text{ubm}} \odot \frac{\frac{\mat{V}_{\text{all}}}{\hat{\mat{V}}_{\text{all}}} \mat{H}_{\text{ubm}}\tr}{\mathbf{1}_{d \times t} \mat{H}_{\text{ubm}}\tr},  \label{eq:Wubmupdate}\\
\mat{H}_{\text{ubm}} & \leftarrow \mat{H}_{\text{ubm}} \odot \frac{ \mat{W}_{\text{ubm}}\tr{ \frac{\mat{V}_{\text{all}}}{\hat{\mat{V}}_{\text{all}}}}}{ \mat{W}_{\text{ubm}}\tr \mathbf{1}_{d \times t} + \lambda} \label{eq:Hubmupdate},
\end{align}
\end{subequations}
where $\odot$ means element-wise multiplication and the division is element-wise.

We stack the columns $\mat{W}_{\text{ubm}}$ to form the UBM dictionary supervector $\vec{w}_{\text{ubm}}$. This will act as the mean supervector for the rest of the steps. 

\subsection{Step 2: Calculation of Sufficient Statistics and Total Variability Matrix}
\label{sec:tvm}

In this step we calculate the 0th and 1st order sufficient statistics from each of the training files and use that to estimate the total variability matrix. We assume that the magnitude spectrograms $\mat{V}$ are drawn from a multivariate log-normal distribution, so $\log \left( \mat{V} \right)$ is drawn from a multivariate normal distribution. Thus, we will calculate the statistics for and model the total variability subspace of $\log \left( \mat{V} \right)$. Note that all the matrices involved have non-negative entries, so there are no issues when taking the log.

For each utterance $u$ in the training set, we calculate it's magnitude spectrogram $\mat{V}_u \in \mathbb{R}_+^{d \times t_u}$. Using $\mat{W}_{\text{ubm}}$ as calculated in Step 1, we find the $k \times t_u$ activation matrix $\mat{H}_u$ as:
\begin{dmath}
\label{eq:input_timaactivation}
\mat{H}_u \leftarrow \mat{H}_u \odot \frac{ \mat{W}_{\text{ubm}}\tr { \frac{\mat{V}_u}{\hat{\mat{V}}_u} }}{ \mat{W}_{\text{ubm}}\tr \mathbf{1}_{d \times t} + \lambda} ,
\end{dmath}
where $ \hat{\mat{V}}_u \vcentcolon =  \mat{W}_{\text{ubm}} {\mat{H}}_u$. Define $\tilde{\mat{H}}_u$ as the columns of $\mat{H}_u$ normalized to sum to $1$.
Then, each column of $\tilde{\mat{H}}_u$ represents the probability distribution of each time frame of $\mat{V}_u$ being represented by the vectors in the dictionary. 

Next, we calculate the $0^{th}$-order statistic $\vec{n}_u$, $1^{st}$-order statistic $\mat{F}_u$, and centered $1^{st}$-order statistic $\bar{\mat{F}}_u$ for each training utterance $u$ as:
\begin{subequations}
\label{eq:ss}
\begin{align}
\mathbf{\vec{n}}_u & = \sum_{\tau = 1}^{t_u} \tilde{\mat{H}}_{u (:, \tau)}  \label{eq:N}\\
\mat{F}_u & = \log \left( \mat{V}_u  \left( \tilde{\mat{H}}_u \mat{N}_u^{-1} \right) \tr  \right) \mat{N}_u \label{eq:F} \\
\bar{\mat{F}}_{u \left( :, c \right)} & = \mat{F}_{u \left( :, c \right)} - n_{uc} \log \left( \mat{W}_{\text{ubm} \left( :, c \right)} \right) \; \forall c \in \left\{ 1, \dots, k \right\}
\end{align}
\end{subequations}
where $\mat{N}_u \vcentcolon = \operatorname{diag} \left( \vec{n}_u \right)$ is a diagonal matrix formed by $\vec{n}_u$. Notice that $\vec{n}_u$ is a vector of size $k \times 1$ and $\mat{F}_u$ is a matrix of size $ d \times k $, where $d$ is the dimension of the features.

We estimate the covariance matrix $\mathbf{\Sigma}_c$ as below:
\begin{multline}
 \mathbf{\Sigma}_c = \frac{1}{n_{uc}} \sum_{u=1}^{U} \tilde{\mat{H}}_{u (c, t)} \left( \log \left( \mat{V}_{u (:,t)} \right) - \log \left( \mat{W}_{\text{ubm} (:,c)} \right) \right) \\
 \quad\quad\quad \left( \log \left( \mat{V}_{u (:,t)} \right) - \log \left( \mat{W}_{\text{ubm} (:,c)} \right) \right) \tr,
\end{multline}
for $c \in \{1, \dots, k\}$.
Once we have calculated the sufficient statistics for each input utterance, we use this to estimate the total variability matrix $\mat{T}$ by performing iterations of the EM algorithm. In each iteration, the posterior mean and covariance of the ivectors $\vec{q}_u$ are estimated during the E step given the current estimate of $\mat{T}$ as below:
\begin{subequations}
\label{eq:ivector}
\begin{align}
\vec{\mu}_{\vec{q}_u} &= \left( \mat{I} + \mat{T} \tr \mat{\Sigma}^{-1} \mat{N}_u \mat{T}\right)^{-1} \mat{T} \tr \mat{\Sigma}^{-1} \bar{\vec{f}}_u \\
\mat{\Sigma}_{\vec{q}_u} &= \left( \mat{I} + \mat{T} \tr \mat{\Sigma}^{-1} \mat{N}_u \mat{T} \right)^{-1}
\end{align}
\end{subequations}
where $\bar{\vec{f}}_u$ is a supervector formed by stacking together columns of the matrix $\bar{\mat{F}}_u$ and $\mat{\Sigma}$ is the block-diagonal covariance matrix formed by $\mat{\Sigma}_c \; \forall c \in \left\{ 1, \dots, k \right\}$. Then, in the M step, the matrices $\mat{T}_c$ are updated as below:
\begin{gather}
\label{eq:T_matrix}
\mat{T}_c \leftarrow \left( \sum_{u=1}^{U} \bar{\mat{F}}_{u(:,c)} \vec{\mu}_{\vec{q}_u}\tr \right) \left(\sum_{u=1}^{U} n_{uc} \left( \mat{\Sigma}_{\vec{q}_u} + \vec{\mu}_{\vec{q}_u} \vec{\mu}_{\vec{q}_u} \tr \right) \right)^{-1}
\end{gather}
for $c \in \{1, \dots, k\}$.

\subsection{Step 3: Extraction of Features}
\label{sec:adapt}

Given an utterance $a$, we find it's magnitude spectrogram $\mat{V}_a$ and use this to calculate $\mat{H}_a$, $\tilde{\mat{H}}_a$, $\mathbf{n}_a$, and $\mat{F}_a$ using Equations~\ref{eq:input_timaactivation} and  \ref{eq:ss}. Then, given $\vec{n}_a$, $\mat{F}_a$, $\mat{T}$, $\vec{w}_{\text{ubm}}$, and $\mat{\Sigma}$, we obtain the i-vector $\vec{q}_a$ for this utterance using the posterior mean estimate given in Equation~\eqref{eq:ivector}.

Thereafter, we calculate the adapted dictionary supervector (which now models the noise type and microphone factors because this information was captured by $\vec{q}_a$) as:
\begin{dmath}
\label{eq:adapted}
\vec{w}_a = \exp \left( \log \left( \vec{w}_{\text{ubm}} \right) + \mat{T} \vec{q}_a \right)  
\end{dmath}
Notice that $\vec{w}_a$ a $kd \times 1$ vector, so we reshape $\vec{w}_a$ to get a $d \times k$ non-negative adapted dictionary $\mat{W}_a$. With $\mat{W}_a$ fixed, we run the NMF algorithm on $\mat{V}_a$ to find the corresponding activation matrix $\mat{H}_a$. 
We use $\log \left( \mat{H}_a \right)$ as features for training the acoustic model in an ASR system.


\section{Experiments and Results}
\label{sec:experiments}

We investigated the performance of our algorithm on the clean speech in the Aurora 4 corpus \cite{aurora4} with added noise from the DEMAND dataset \cite{demand}. The training set consists of 7138 utterances from the Aurora 4 training set corrupted by one of six different noises (labeled in the DEMAND dataset as ``dliving'', ``npark'', ``omeeting'', ``presto'', ``straffic'', and ``tcar'') at 5--15~dB SNR. The test set consists of 330 utterances from 8 speakers, with each of the utterances corrupted by the same six noises with SNRs ranging from 5--15~dB. Additionally, we created a second test set with ``ohallway'', ``pstation'', and ``spsquare'' noises added to test the ASR performance in unseen noise conditions.

We compared our proposed features to MFCCS, PNCCs \cite{pncc}, and the noise-robust features proposed in \cite{vaz} (we will refer to these features as CNMF Features since that algorithm uses convolutive NMF (CNMF) to generate features). For our proposed features, we performed NMF on $d = 129$-dimensional spectrograms with $k = 60$ dictionary vectors. We set the dimension of the total variability subspace $s = 400$. To match the parameters for the proposed features, we also used $60$ dictionary vectors for the CNMF Features. We extracted 13-dimensional MFCC and PNCC features. For each of the features, we applied a speaker-independent global mean and variance normalization prior to augmenting them with delta and delta-delta, followed by Linear Discriminant Analysis (LDA) and Maximum Linear Likelihood Transform (MLLT). We input the transformed features into a fully-connected 4-layer neural network, with 1024 hidden nodes per layer. The network uses tanh non-linearities and minimizes the cross-entropy using stochastic gradient descent.

Table~\ref{tab:results} shows the word error rates (WER) on the Aurora 4 + DEMAND test set for our proposed features and the three baseline features for clean speech, noisy speech with noise seen during training, and noisy speech with noise not seen during training. We also provide the weighted average WER for the three conditions. From the results, one can see that clean speech has the lowest WER while the performance degrades with noise for all feature sets. At first glance, it seems surprising that the performance on unseen noises is better than on seen noises. But we note that in general the seen noises more background speech and non-stationary characteristics compared to the unseen noises, and these characteristics, particularly background speech, makes ASR more challenging. Indeed, when we inspected the results, we noticed that performance on ``presto'' noise (restaurant noise) had more than twice the WER compared to other noises.

\begin{table}[h]
    \centering
    \caption{WER for the Aurora 4 + DEMAND test set in the three noise conditions.}
    \label{tab:results}
    \begin{tabular}{l|c|c|c|c} 
        \hline
        \hline
        Condition & MFCC & PNCC & CNMF & Proposed \\
        \hline
        clean & $3.68$ & $3.96$ & $3.40$ & $5.60$ \\
        noisy, seen & $6.05$ & $6.71$ & $6.04$ & $8.53$ \\
        noisy, unseen & $4.27$ & $4.64$ & $4.38$ & $5.95$ \\
        \hline
        Average & $5.28$ & $5.81$ & $5.28$ & $7.46$ \\
        \hline
        \hline
    \end{tabular}
\end{table}


Unfortunately we were not able achieve a lower WER with the proposed features compared to the baseline features. This is most likely due to the fact that the proposed features are computed from a per-utterance transform while MFCC, PNCC, and CNMF features are computed from fixed transforms, so the proposed features are sensitive to the parameters chosen when learning the transform and how well the total variability matrix $\mat{T}$ captures the sources of variability in the training set. In particular, the subspace dimension $s$ is very important because underestimating $s$ leads to poor modeling of the sources of variability, while overestimate $s$ will result in $\mat{T}$ capturing extraneous information. Moreover, we believe that $7138$ utterances in the training set may not be an adequate amount of observations to properly learn the total variability subspace. In fact, our initial experiments were carried out the standard Aurora 4 dataset, which includes microphone variability in addition to speaker and noise variability. Given the limited amount of training data, having additional variability due to channel conditions resulted WER that was much greater for Aurora 4 than for Aurora 4 + DEMAND\footnote{The WER for all feature sets was greater with Aurora 4 compared to Aurora 4 + DEMAND due to the multicondition style training. However, the amount of change in WER was much greater for the proposed features compared to the baseline features}. Therefore, we are confident that a larger training set should result in a WER competitive with the other features as it will allow for better modeling of the total variability subspace.

On the other hand, one can see that the proposed features have the smallest gap in WER between the clean and unseen noise conditions compared to the baseline features. The main motivation behind this work is to generate acoustic features that are robust to acoustic conditions, so this result gives an indication that our algorithm can achieve this goal. This finding alone gives us good reason to improve the overall performance because it will allow ASR systems to perform near clean speech WER without requiring to re-train the acoustic model for specific acoustic conditions.


\section{Conclusion}
\label{sec:conclusion}

We proposed an algorithm to calculate noise-robust acoustic features from noisy utterances. The algorithm uses Total Variability Modeling to learn a total variability subspace and adapt a UBM NMF dictionary for each utterance at test time. We use the NMF activation matrix corresponding to the adapted dictionary as the acoustic features. Thus, our proposed features are calculated from per-utterance transforms, which could lead to greater robustness to the specific noise present in each utterance. Moreover, our algorithm, which builds upon the work in \cite{vaz}, does not require a training dataset of parallel clean and noisy speech. While the proposed features did not perform better than baseline features on the Aurora 4 + DEMAND corpus, we note that the WER was more consistent across clean and noisy conditions, in particular the unseen noise condition. This gives an indication that our approach has the potential to perform robustly in different noise conditions, but the overall results suggest that we should explore training with a larger dataset to better learn the total variability subspace.

Going forward, we will first retrain our system on a larger dataset, such as the Librispeech corpus \cite{librispeech}. Also, we will test the performance of the proposed algorithm in the presence of channel variability to more accurately simulate real-world acoustic conditions.


\bibliographystyle{IEEEtran}
\bibliography{refs}

\end{document}